\documentclass{aa}
\usepackage{graphicx}
\def\ltsim{\raise 2pt \hbox {$<$} \kern-1.1em \lower 4pt \hbox {$\sim$}}
\def\ltapprox{\raise 2pt \hbox {$<$} \kern-1.1em \lower 5pt \hbox {$\approx$}}
\def\gtsim{\raise 2pt \hbox {$>$} \kern-1.1em \lower 4pt \hbox {$\sim$}}
\def\gtapprox{\raise 2pt \hbox {$>$} \kern-1.1em \lower 5pt \hbox {$\approx$}}

\def\arcsec{$^{\prime\prime}$}
\def\arcmin{$^{\prime}$}
\def\degrees{$^{\circ}$}
\def\etal{{\rm et al.~}}
\begin{document}
\title
{Deep images of cluster radio halos}
\author{M. Bacchi\inst{1,2} 
\and L. Feretti\inst{2} 
\and G. Giovannini\inst{2,3} 
\and F. Govoni\inst{2,3}}
\offprints{L. Feretti}
\institute{
 Dipartimento di Fisica, Univ. Bologna,
Via B. Pichat 6/2, I--40127 Bologna, Italy
\and
Istituto di Radioastronomia -- CNR, via Gobetti 101, I--40129
Bologna, Italy
\and Dipartimento di Astronomia, Univ. Bologna,  
Via Ranzani 1, I--40127 Bologna, Italy}

   \date{Received ... ; accepted ...}

\abstract{
New radio data are presented for the clusters A401, A545,
A754, A1914, A2219 and A2390, where the presence of 
diffuse radio emission was suggested from the images of the
NRAO VLA Sky Survey. Sensitive images of these clusters, 
obtained with the Very Large Array (VLA)
at 20 cm  confirm the existence of the diffuse sources and allow 
us to derive their fluxes and intrinsic parameters. 
The correlation between the halo radio power and cluster X-ray
luminosity is derived for a large sample of halo clusters, and
is briefly discussed.
\keywords{Radio continuum: general - Galaxies: clusters: general - 
Intergalactic medium - X-rays: galaxies: clusters }
}

\maketitle

\section{Introduction}

Great attention has been devoted in recent years to the study of
cluster large-scale diffuse radio sources that have no obvious
connection to any individual cluster galaxy.  The sources classified
as {\it radio halos} are generally regular in shape and located at the
cluster centers.  Sources with similar properties have also been found
at the cluster peripheries: they are the {\it relic} sources.  They
commonly show irregular and elongated shape and exhibit stronger
polarization than halos.  Moreover, in some clusters with a central
powerful radio galaxy, the relativistic particles can be traced out to
distances up to $\sim$500 kpc, forming what is called a {\it
mini-halo} (see e.g. 3C~84 in the Perseus cluster, Burns \etal 1992).
These sources demonstrate the existence of relativistic electrons and
large scale magnetic fields in the intracluster medium, so they probe
that non-thermal processes are common in clusters of galaxies.

The difficulty in explaining these diffuse sources arises from the
combination of their large size and the short lifetimes of radiating
electrons.  Two main classes of theoretical models have been proposed
for the origin of radio halos: (i) in-situ electron reacceleration by
turbulent gas motion or shocks produced in the intergalactic medium
during cluster merger processes; (ii) secondary particle production by
hadronic interaction of relativistic protons with the background gas
of the IGM.  Also the relics have been suggested to be connected to
shocks in cluster merger processes, either through diffusive shock
acceleration of radio emitting particles (En{\ss}lin \etal 1998) or
adiabatic recompression of fossil radio plasma (En{\ss}lin \&
Br\"uggen 2002).  On the contrary, the mini-halos are not linked to
cluster merger processes.  A recent models involving electron
reacceleration due to MDH turbulence in the cooling flow has been
suggested by Gitti \etal (2002a).

The knowledge of the physical properties of these sources, and of
their origin and evolution is limited by the low number of halos and
relics well studied up to now.  Also, it is not yet clear if radio
halos and relics have a common origin and evolution, or should be
considered as different classes of sources.

To properly map these extended sources one needs large sensitivity to
the extended features, but also high resolution to
distinguish a real halo from the blend of unrelated sources.

\begin{table*}
\caption{Cluster properties}
\begin{flushleft}
\begin{tabular}{llllllllll}
\hline
\noalign{\smallskip}
Name & z &  RA \ (2000) &   DEC & BM  & kpc/\arcsec  & T & L$_{\rm X~bol}$ \\
     &   & h\ \  m\ \  s & \ \degrees \ \ \ \arcmin \ \ \ \arcsec & 
& & keV & erg s$^{-1}$ \\
\noalign{\smallskip}
\hline
\noalign{\smallskip}
A401 & 0.0737  &  02 58 58.3 &  +13 34 36  &  I   & 1.89 & 8.0 & 2.54E+45  \\
A545 & 0.154   &  05 32 25.3 & --11 32 35  & III  & 3.48 & 5.5 & 2.11E+45 \\
A754 & 0.0542  &  09 09 20.6 & --09 40 35  & I-II & 1.44 & 9.0 & 2.22E+45 \\
A1914 & 0.1712 &  14 26 01.1 &  +37 49 26  & II   & 3.77 & 10.5 & 5.63E+45 \\
A2219 & 0.2256 &  16 40 21.1 &  +46 41 59  & III  & 4.59 & 12.4 & 6.29E+45 \\
A2390 & 0.228  &  21 53 37   &  +17 41 41  &  I   & 4.62 & 11.1 & 6.05E+45 \\
\noalign{\smallskip}
\hline
\noalign{\smallskip}
\end{tabular}
\end{flushleft}
Caption. Col. 1: cluster name; Col. 2: redshift from Struble \& Rood
(1999);
Cols. 3 and 4: coordinates of the X-ray cluster center from 
David \etal (1999), but A2390 from B\"ohringer \etal (1998);
Col. 5: Bautz-Morgan type; Col. 6: linear to angular conversion;
Col. 7: temperature from Wu \etal (1999) but A1914 from
White (2000);
Col. 8: bolometric X-ray luminosity from a revised version of the table
given in Wu \etal (1999).  
\end{table*}

We present here sensitive images of 6 clusters of galaxies, where the
presence of diffuse sources was suggested by Giovannini \etal (1999)
from a search in the NRAO\footnote {The National Radio Astronomy
Observatory is operated by Associated Universities, Inc., under
contract with the National Science Foundation.}  VLA Sky Survey (NVSS,
Condon \etal 1998).  As a cluster sample, Giovannini \etal (1999) used
the X-ray brightest Abell clusters reported by Ebeling \etal (1996).
They found 29 clusters, 11 of which were already known in the
literature to contain a radio halo or relic, and 18 were new
detections of cluster diffuse sources.  Images with improved
sensitivity and resolution of 6 clusters (A115, A520, A773, A1664,
A2254 and A2744) have already been presented by Govoni \etal (2001).

The observations of the 6 clusters A401,
A545, A754, A1914, A2219 and A2390, were carried out using the Very Large
Array (VLA) in the C and D configurations at 20 cm.  For each extended
source, we derived the total flux densities, 
the size, and the physical conditions.

For the computation of intrinsic parameters, a Hubble constant
H$_0$=50 km s$^{-1}$ Mpc$^{-1}$ and a deceleration parameter q$_0$=0.5
are adopted.

\section{Radio Observations}

The general properties of the clusters under study are given in Table
1.  The choice of the observing
frequency and of the configurations ensures a good sampling of short
spacings, to adequately image the extended diffuse sources.  The
shortest baseline is 35 m,  thus
allowing to  detect structures up to about $\sim$15\arcmin~ in angular
size.  The details of the radio observations are given in Table 2.
Data were calibrated and reduced with the Astronomical Image
Processing System (AIPS), following the standard procedure (Fourier
inversion, CLEAN and RESTORE, Self-Calibration).  Significant editing
of the data was needed because of strong interferences, affecting in
particular the D array observations in the higher frequency IF. In
some cases, this full data set was omitted. Therefore the noise in the
final maps is much higher than the theoretical thermal noise.

\begin{table}
\caption{Observing log}
\begin{flushleft}
\begin{tabular}{lllllr}
\hline
\noalign{\smallskip}
Name & Freq. &  Bw  & Array  &  Date  & Durat. \\
  & MHz & MHz  & & & min. \\
\noalign{\smallskip}
\hline
\noalign{\smallskip}
A401  & 1365/1435 & 50 & D &  SEP 2000 & 160 \\
A545  &  1365/1435 & 50 & C &  APR 2000 & 130 \\ 
       &  1365/1435 & 50 & D &  SEP 2000 & 160 \\
A754  &  1365/1435 & 50 & D &  SEP 2000 & 160 \\
A1914 & 1365/1435 & 50 & C &  APR 2000 & 160 \\
       & 1365/1435 & 50 & D &  SEP 2000 & 160 \\
A2219 & 1365/1435 & 50 & C &  APR 2000 & 160 \\
       & 1365/1435 & 50 & D &  SEP 2000 & 160 \\
A2390 & 1365/1435 & 50 & C &  APR 2000 & 60 \\
\noalign{\smallskip}
\hline
\noalign{\smallskip}
\end{tabular}
\end{flushleft}
\end{table}

Data from the two different configurations were first reduced
separately, for a consistency check, and then added together to
produce the final images.  Images with angular resolutions ranging
from $\sim$15\arcsec~ to 70\arcsec~ were produced.
We also obtained maps of the polarized intensity in the standard way.

\section{Results}

We give below the description of the results for each cluster and the
images of the diffuse radio emission detected in all of them.  We
summarize in Table 3 the properties of the diffuse sources. The total
flux densitiy refers to the diffuse emission, after subtraction of
discrete sources. The volume was evaluated by assuming an ellipsoidal
shape, with the depth equal to the minimum size. The minimum energy
density and equipartition magnetic field were computed using standard
formulae (Pacholczyk 1970), integrating the radio emission between 10
MHz and 10 GHz, with a radio spectral index $\alpha$ = 1.2
\footnote {S($\nu) \propto \nu^{-\alpha}$}, and
assuming equal energy density in protons and electrons, and a volume
filling factor of 1.

\begin{figure}
\includegraphics{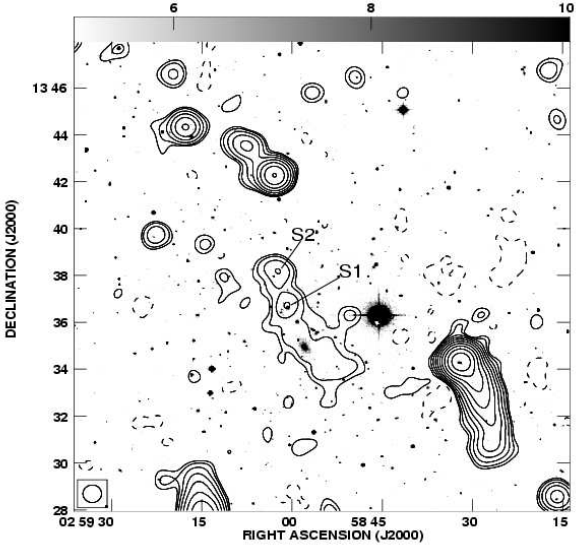}
\vspace{8 cm}
\caption{Contour map of the radio emission 
at 20 cm in A401, overlaid onto the Digital Sky Survey optical
grey-scale.
The resolution is 45\arcsec$\times$45\arcsec. The noise level is 
0.13 mJy/beam.
Contour levels are -0.3, 0.3, 0.5, 1, 2, 4, 8, 16, 32, 64, 128  mJy/beam.
Labels indicate discrete sources.
}
\end{figure}

\subsection{Abell 401} 

The presence of diffuse emission in this cluster, located north of the
central cD galaxy, was reported by Harris \etal (1980a) and Roland
\etal (1981). The last authors  identified the possible radio halo
with their source 14W19-b and reported a flux density of 80 mJy at
610 MHz, with a spectral index $\alpha \sim$1.4.
Unlike the previous images, a faint radio emission located around
the central cD galaxy was detected in the NVSS image (Giovannini \etal
1999), with a total flux density at 1.4 GHz of 21 mJy. 

The deep radio map of this cluster, shown in Fig. 1, confirms the
presence of a diffuse radio source around the cD galaxy, as found in
the NVSS. The total flux density in the present image
is lower than that 
reported by Giovannini \etal (1999), because of the accurate
subtraction of the discrete sources.
The diffuse emission is not detected in the short
observation  obtained with the VLA at 330 MHz 
by Giovannini \& Feretti (2000).  
As discussed by  Kassim \etal (2001) for the cluster A754,
the extended low brightness emission could be
missed at 330 MHz because of the poor uv coverage. Therefore, 
the lack of detection at 330 MHz cannot be used to 
obtain an upper limit to the radio spectral index $\alpha^{1.4}_{0.3}$.

The two emission peaks S1 and S2 have been considered as two possible
unrelated sources.  Thus their flux has been subtracted in
the computation of the total flux of the halo.  The classification of
this diffuse source as a radio halo is only based on the fact that it
is located at the cluster center. However, unlike typical radio halos,
its shape is very elongated and irregular, more similar to peripheral
radio relics.

No polarization is detected down to a limit of $\sim$4\%.
This fact would not support the possibility that this is a relic 
projected onto the cluster center.

According to the ASCA temperature map, Markevitch \etal (1998) 
suggested that
this cluster is beginning a collision with the near cluster A399. They
considered A401 as a rather unusual cluster, since neither temperature
map nor the X-ray image indicate recent merger activity in the central
cluster region, and yet it has no cooling flow (White \etal 1997,
Peres \etal 1998).  On the other hand, from the wavelet analysis of
the ROSAT X-ray brightness distribution, Slezak \etal (1994)
determined a complex structure and obtained a wavelet image which, we
note, is remarkably similar to the structure of the radio halo.

\begin{figure}
\includegraphics{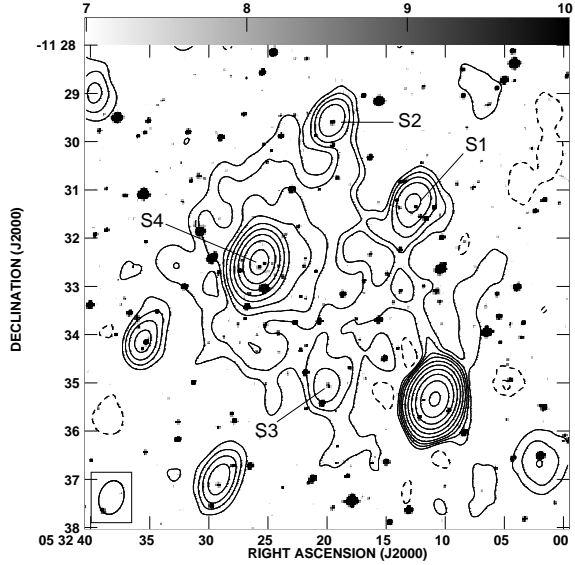}
\vspace{8 cm}
\caption{Contour map of the radio emission 
at 20 cm in A545, overlaid onto the Digital Sky Survey optical
grey-scale.
The resolution is 42\arcsec$\times$30\arcsec (at PA=--17\degrees). 
The noise level is 0.045 mJy/beam.
Contour levels are -0.175, 0.175, 0.35, 0.6, 1, 2, 4, 8, 16, 32, 64  mJy/beam.
Labels indicate discrete sources.
}
\end{figure}

\subsection{Abell 545} 

The radio image of this cluster is presented in Fig. 2, where the
extended radio halo is easily visible. Its structure is rather regular
and centrally located.

A strong discrete source is present within the radio halo at
RA$_{2000}$ = 05$^h$ 32$^m$ 25.7$^s$, DEC$_{2000}$ = --11\degrees~
32\arcmin~ 30.2\arcsec (source S4 in Fig. 2), i.e,
within 8.5\arcsec~ from 
the position of the brightest cluster galaxy given by Schneider \etal
(1983). The identification between the radio source and the
galaxy cannot be established with the present  data but it is
unlikely. Indeed, no cluster radio galaxies are reported in
this cluster by Ledlow \& Owen (1995).
Another strong radio source, located at the south-western 
halo boundary, is clearly unrelated to the diffuse radio emission.
 No higher resolution image  of this region  is available
from the literature or the VLA Faint Images of the Radio Sky at
Twenty-cm Survey (FIRST, Becker \etal 1995).

The total flux density reported by Giovannini \etal (1999) from the
NVSS image is 41 mJy. This is higher than the present flux density
estimate, due to the accurate subtraction here of discrete sources.
For this cluster, a comparison between the data of the two separate
frequencies 1365 and 1435 MHz allowed to derive that the spectrum is
steep with a lower limit $\alpha>$1.4.  The central part of the halo
is not polarized at a level of a few percent. At the boundaries, the
upper limit to the polarized flux is, however of $\sim$15\%.

The X-ray structure of this cluster is highly elongated within 0.5
Mpc, which may reflect dynamical youth of the interior (Buote \& Tsai
1996, Buote 2001).  Its unrelaxed state is consistent with the absence
of a cooling flow (White \etal 1997).

\begin{table*}
\caption{Properties of the radio halos}
\begin{flushleft}
\begin{tabular}{lllllllll}
\hline
\noalign{\smallskip}
Name & S$_{1.4}$ & $\theta_{\rm max} \times \theta_{\rm min}$ & LLS & Volume & P$_{1.4}$ & u$_{\rm min}$ & 
H$_{\rm eq}$ & Type \\
     &          &                                    &      &  $\times$10$^7$
&  $\times$ 10$^{24}$ &  $\times$ 10$^{-14}$ \\
      &   mJy    & \arcmin \ \ \ \ \ \ \ \ \arcmin & kpc   &  kpc$^3$ 
&  W Hz$^{-1}$ &  erg cm$^{-3}$  & $\mu$G &     \\
\noalign{\smallskip}
\hline
\noalign{\smallskip}
A401  & 17$\pm$1  & 6.3 $\times$ 2    & 700  & 1.7 & 0.41 & 4.5  & 0.7 & H \\
A545  & 23$\pm$1  & 5.6 $\times$ 5.3  & 1150 & 75  & 2.52 & 1.4  & 0.4 & H \\
A754  & 86$\pm$4  & 15.8 $\times$ 4.6 & 1350 & 11  & 1.12 & 1.8  & 0.6 & 
H \\
       & 69$\pm$3  & 13 $\times$ 10    & 1130 & 44  & 0.89 & 1.1  & 0.3 & 
R \\
A1914 & 64$\pm$3  & 7.4 $\times$ 5.3  & 1650 & 120 & 8.72 & 2.3  & 0.5 & H \\
A2219 & 81$\pm$4  & 8 $\times$ 6.3    & 2180 & 340 & 19.6 & 2.0  & 0.5 & H \\
A2390 & 63$\pm$3  & 2 $\times$ 1.7    & 550  & 6.5 & 15.6 & 16.8 & 1.3 & MH \\
\noalign{\smallskip}
\hline
\noalign{\smallskip}
\end{tabular}
\end{flushleft}
Caption. Col. 1: cluster name; Col. 2: flux at 1.4 GHz;
Cols. 3: angular size (maximum $\times$ minimum); Col. 4: maximum linear size;
Col. 5: source volume; Col. 6: radio power at 1.4 GHz; 
Col. 7: minimum energy density; Col. 8: equipartition magnetic
field; Col. 9: type of diffuse source: H = halo, R = relic, MH = mini-halo.
\end{table*}

\begin{figure}
\includegraphics{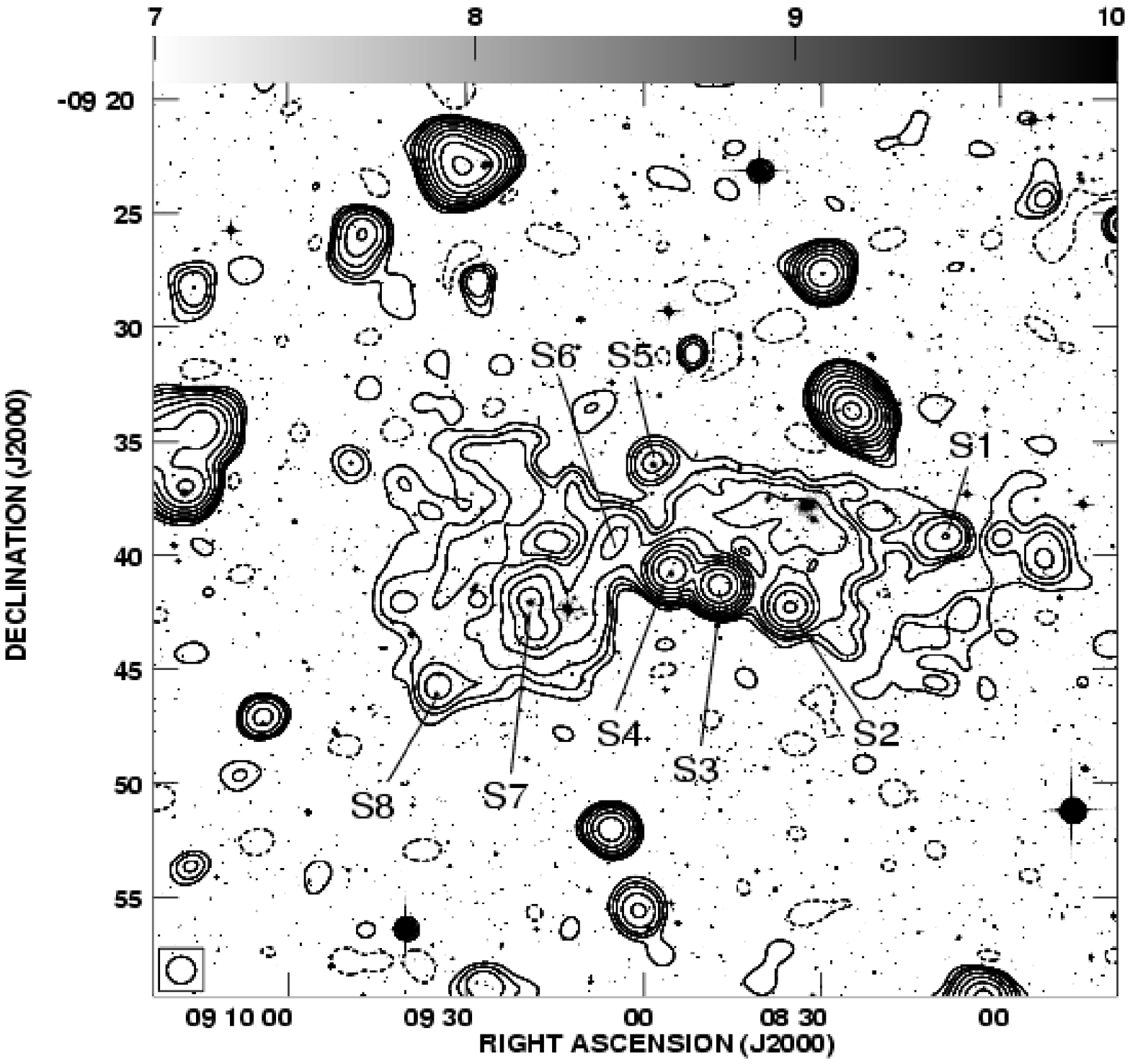}
\vspace{8 cm}
\caption{Contour map of the radio emission 
at 20 cm in A754, overlaid onto the Digital Sky Survey optical
grey-scale.
The resolution is 70\arcsec$\times$70\arcsec. 
The noise level is 0.1 mJy/beam.
Contour levels are -0.3, 0.3, 0.6, 1, 2, 4, 8, 16, 32, 64, 128  mJy/beam.
Labels indicate discrete sources.
}
\end{figure}

\begin{figure}
\includegraphics{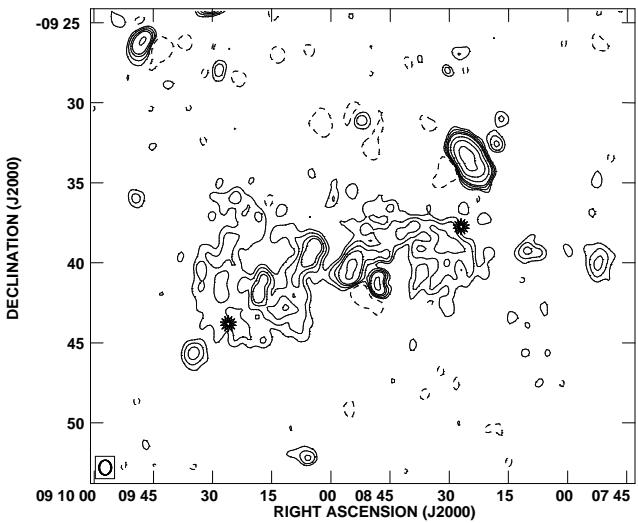}
\vspace{8 cm}
\caption{Contour map of the radio emission 
at 20 cm in A754, after subtraction of the discrete sources.
The resolution is 55\arcsec$\times$45\arcsec (DEC$\times$RA). 
The noise level is 0.13 mJy/beam.
Contour levels are -0.4, 0.4, 0.8, 1.4, 2, 4, 8, 16  mJy/beam.
The star-symbols indicate the centers of the optical cluster clumps.}
\end{figure}

\subsection{Abell 754} 

The existence of a halo in this cluster was suggested by Harris \etal
(1980b) and recently confirmed by the image at 0.3 GHz of Kassim \etal
(2001). The image obtained with the present data is shown in Fig. 3,
where the resolution has been slightly degraded, to improve the signal
to noise ratio.  The diffuse emission detected here is complex and
very extended, consisting of two main features.  A diffuse source is
visible around the brightest optical galaxy, classified as cD galaxy
by Dressler (1980).  Moreover, diffuse emission is detected to the
east, consistent with the NVSS image (Giovannini \etal 1999) and with
the 74 MHz image of Kassim \etal (2001). This feature is classified 
by the latter authors as the east relic. 
We note that in Fig. 3, there are holes of radio emission
north of the discrete sources S7 and S3. 
We do not detect in our image any
radio emission in the region of the western relic of Kassim \etal
(2001).  

A possible connection between the two components of the diffuse
emission cannot be excluded on the basis of our image.  To better
investigate this point, we produced an image where the discrete
sources embedded within the diffuse emission are subtracted.  The
procedure was as follows: we first obtained an image of discrete
sources, by selecting from the original data set only the data from
the long baselines, which do not contain the extended halo
emission. This image was then subtracted from the total image. The map
obtained in this way is presented in Fig. 4, which has a higher
resolution with respect to that reported in Fig. 3, to better show the
structure.  It is clear from Fig. 4 that the subtraction of discrete
sources is only partial. We note, however, that the two diffuse
sources are both irregular in shape, and there is a deficit of radio
emission in between them. Therefore we consider them as two separate
sources. The diffuse emission west of sources S1 and S2 in Fig. 3 is
not visible here, because of its very low surface brightness.

The polarization image is quite noisy, allowing only to derive upper
limits of 9\% and 15\% to the polarization degree of the western and
eastern diffuse features, respectively.

The determination of the center of this cluster is not obvious, since
the cluster shows a very asymmetric morphology in X-ray and two major
clumps in the optical, neither of them lying on the off-center peak of
the X-ray emission (Zabludoff \& Zaritsky 1995). The center given in
Table 1 is close to the X-ray peak, which is about 3.5\arcmin~ north
of the eastern optical peak.  In Fig. 4, we indicate with star-symbols
the centers of the two optical clumps, as given by Zabludoff \&
Zaritsky (1995). It is interesting to note that the two components of
the diffuse emission are associated with the two optical clumps.  The
classification of these diffuse sources as a halo and a relic is
tentative. It is based on the presence of a cD galaxy in the western
optical clump, and therefore the indication that this clump is the
main cluster.  Alternatively, the two diffuse sources could be i) two
halos according to the optical galaxy distributions, ii) two relics
according to the X-ray image.  We can generally conclude that the
diffuse emission in A754 is quite complex, reflecting the complexity
of the dynamical state of this cluster (see also the structure
analysis by Buote \& Tsai 1996 and the ASCA temperature map by Markevitch
\etal 1998). No cooling flow (White \etal 1997, Peres \etal 1998) is
present in this cluster. Fusco-Femiano \etal (2002)
report the detection of hard X-ray emission at the 3.2$\sigma$ level
from this cluster, possibly of Inverse Compton origin. 
From the comparison between the X-ray data and
the present radio data, they infer a magnetic field
of $\sim$ 0.1 $\mu$G.

For the central radio halo, Kassim \etal (2001) obtain
flux densities of 4 Jy and 750 mJy, at 74 MHz and 330 MHz,
respectively, leading to a spectral index 
$\alpha^{0.3}_{0.07}$ $\sim$ 1.1.
The comparison between the present flux in Table 3 and that at 0.3 GHz
given by Kassim \etal (2001) leads to a spectral index of the halo of
$\alpha^{1.4}_{0.3}$ $\sim$ 1.5.  This is steeper than the spectrum
 in the 74-330 MHz range, indicating the presence of a
possible spectral cutoff.

In the 74 MHz map of Kassim \etal (2001), the relic 
has a flux density of 1.45 Jy. It is significantly
less extended than in the present map and it is located at the eastern
boundary of the diffuse source detected here, including the discrete
source S8.  Therefore, we can only obtain a lower limit to the
spectral index ($\alpha^{1.4}_{0.07}$ $>$ 1). An upper limit
$\alpha^{1.4}_{0.3}$ $<$ 1.7 is derived from the lack of a detection
above 3 sigma in the 330 MHz image.

\begin{figure}
\includegraphics{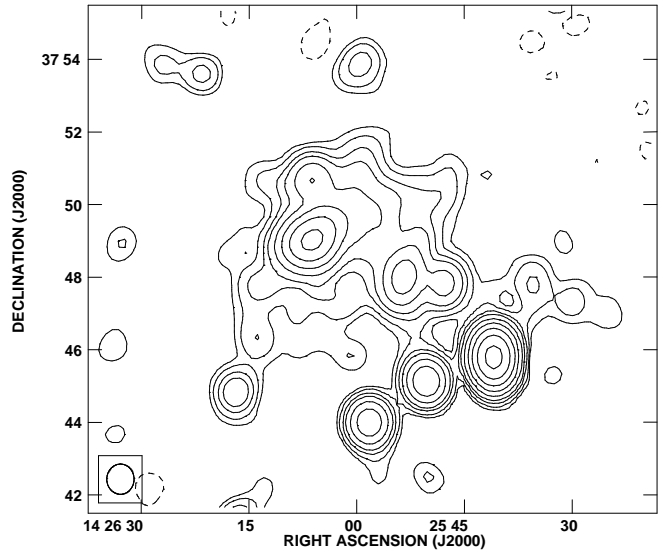}
\vspace{8 cm}
\caption{Contour map of the radio emission 
at 20 cm in A1914, at the resolution of
50\arcsec$\times$45\arcsec (DEC$\times$RA). 
The noise level is 0.05 mJy/beam.
Contour levels are -0.15, 0.15, 0.3, 0.6, 1, 2, 4, 8, 16, 32  mJy/beam.
}
\end{figure}

\begin{figure}
\includegraphics{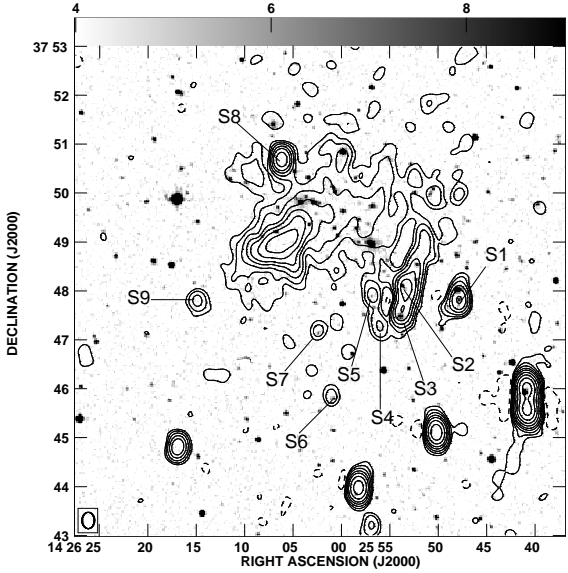}
\vspace{8 cm}
\caption{Contour map of the radio emission 
at 20 cm in A1914, overlaid onto the Digital Sky Survey optical
grey-scale.
The resolution is 20\arcsec$\times$15\arcsec (DEC$\times$RA). 
The noise level is 0.02 mJy/beam.
Contour levels are -0.07, 0.07, 0.14, 0.28, 0.56, 1, 2, 4, 8, 16
  mJy/beam. Labels indicate discrete sources.
}
\end{figure}

\begin{figure}
\includegraphics{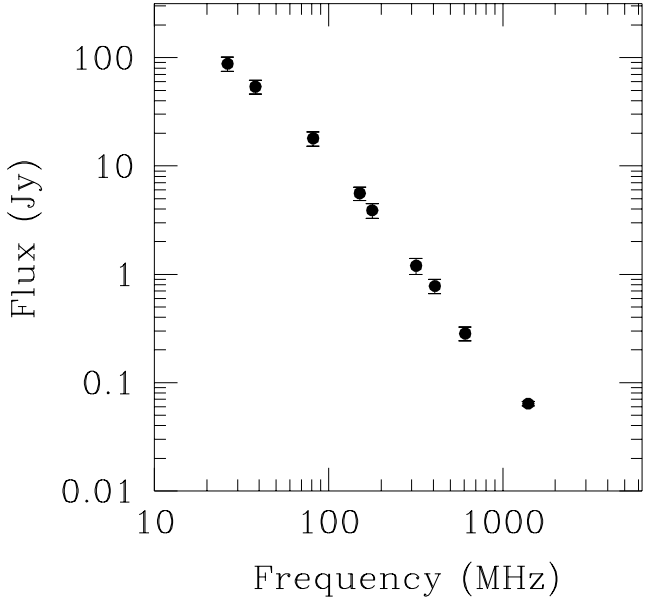}
\vspace{8 cm}
\caption{Integrated spectrum of the diffuse source in A1914
obtained by adding the present flux density at 1.4 GHz to
the data given by  Komissarov \& Gubanov (1994). 
}
\end{figure}

\subsection{Abell 1914}

A very steep spectrum radio source (4C~38.39, $\alpha >$2) was detected
by Komissarov \& Gubanov (1994) and Roland \etal (1985).  The diffuse
emission has been reported also by Kempner \& Sarazin (2001) as a
result of their search for halo and relic candidates in the Westerbork
Northern Sky Survey (WENSS) at 0.3 GHz. The low resolution map
presented in Fig. 5 shows the entire diffuse emission, which is more
complex and extended than in any previous images.
The total flux density estimated here is higher than 
that obtained from the NVSS (50 mJy, Giovannini \etal 1999), 
confirming that some structure is missing in the 
NVSS image. The higher resolution
map in Fig. 6 allows one to distinguish the discrete sources.  The
elongated feature south of source S8 is completely resolved out in
high resolution images (e.g. the FIRST) and therefore is part of the
halo, which looks quite irregular in shape.

A spectrum of this source, using the present data and the flux
densities published by Komissarov \& Gubanov (1994) is given in
Fig. 7. The 0.3 GHz flux of 114 mJy  given by  
Kempner \& Sarazin (2001) is not reported in the
figure, since it is likely 
understimated because of limited sensitivity.
The spectrum is very steep, with an overall slope of $\alpha
\sim$1.8. For a discussion of the spectral behaviour and its possible
curvature we refer to the paper of Komissarov \& Gubanov (1994).

No polarized flux is detected in this halo down to a level of $\sim$
3\%.

In the morphological and
dynamical analysis performed by Buote \& Tsai (1996), the
values of the power ratios P$_1$/P$_0$ and
P$_3$/P$_0$ for this cluster are representative of a disturbed system. 
This is confirmed by the asymmetric X-ray brightness distribution
detected in the Chandra ACIS image, retrieved from
the Chandra archive.  No significant
mass accretion rate is derived by White (2000), indicating that there
is no cooling flow in this cluster.

\begin{figure}
\includegraphics{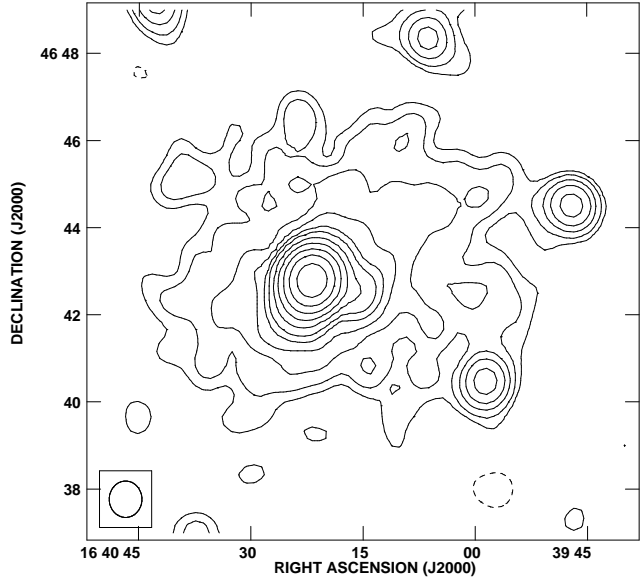}
\vspace{8 cm}
\caption{Contour map of the radio emission 
at 20 cm in A2219, at the resolution of
50\arcsec$\times$45\arcsec (DEC$\times$RA). 
The noise level is 0.07 mJy/beam.
Contour levels are -0.2, 0.2, 0.4, 0.8, 1.4, 2.5, 4, 8, 16, 32, 64
128  mJy/beam.
}
\end{figure}

\begin{figure}
\includegraphics{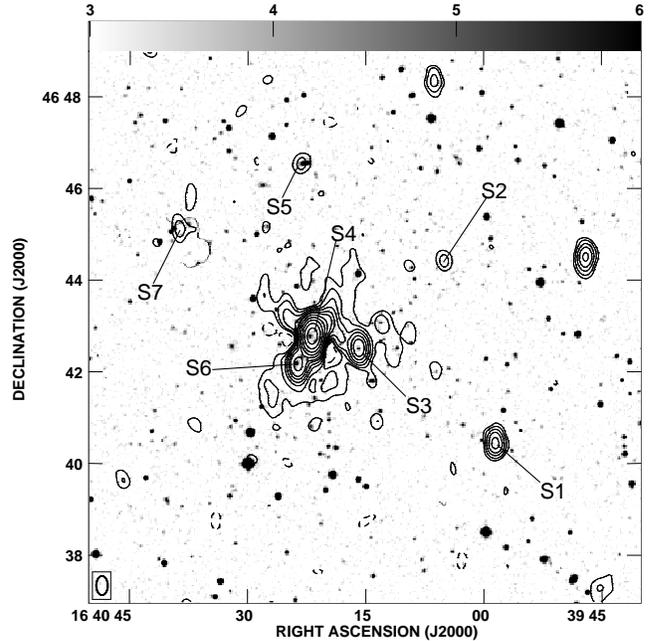}
\vspace{8 cm}
\caption{Contour map of the radio emission 
at 20 cm in A2219, overlaid onto the Digital Sky Survey optical
grey-scale.
The resolution is 25\arcsec$\times$15\arcsec (DEC$\times$RA). 
The noise level is 0.07mJy/beam.
Contour levels are -0.2, 0.2, 0.4, 0.8, 1.4, 2.5, 4, 8, 16, 
32, 64, 128  mJy/beam. Labels indicate discrete sources.
}
\end{figure}

\subsection{Abell 2219}

A giant radio halo is detected in this cluster, as shown in Fig. 8.
The strong radio source at the cluster center is the blend of three
sources (see the high resolution image in Fig. 9) identified with
cluster galaxies (see Fig. 7 and Owen \etal 1992). The northermost
radio galaxy (S4 in Fig. 8) shows a tailed structure (B1638+468, Owen
\& Ledlow 1997).  The polarization map is noisy. Only an upper limit of
$\sim$6.5\% to the polarized flux has been obtained.

The radio halo has a rather regular and symmetric structure,
reminiscent of typical well known halos (Coma, A2255, A2163). It is 10
times more powerful than the halo in the Coma cluster, comparable with
the giant halo in A2163 (Feretti \etal 2001).

Some diffuse emission in this cluster is also detected in the WENSS
image at 0.3 GHz (Kempner \& Sarazin 2001), but their total flux
density is clearly understimated because of limited sensitivity.

From the X-ray image obtained with the Rosat HRI, presented by Rizza
\etal (1998), it is evident that this cluster shows significant
substructure, with two main X-ray peaks displaced by about 1\arcmin.
According to Allen \& Fabian (1998), no cooling flow is present in
this cluster.

\begin{figure}
\includegraphics{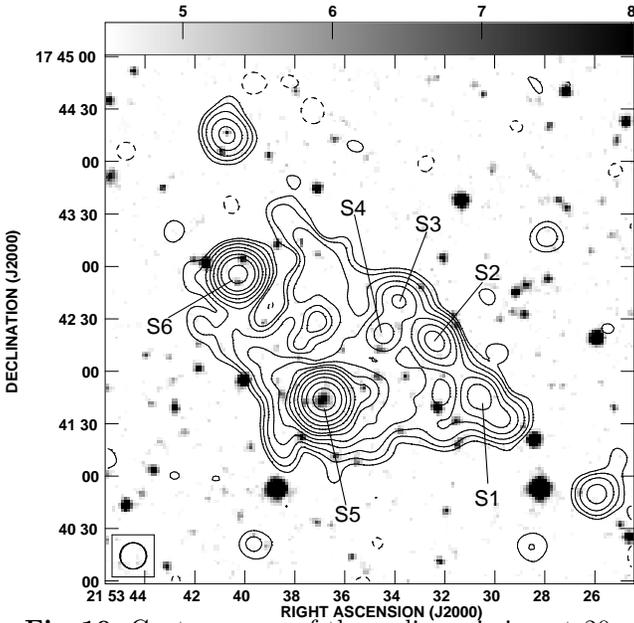}
\vspace{8 cm}
\caption{Contour map of the radio emission 
at 20 cm in A2390, overlaid onto the Digital Sky Survey optical
grey-scale.
The resolution is 15\arcsec$\times$15\arcsec (DEC$\times$RA), 
the noise level is 0.03 mJy/beam.
Contour levels are -0.1, 0.1, 0.2, 0.4, 0.8, 1.4, 2.5, 4, 8, 16, 
32, 64, 128  mJy/beam. Labels indicate discrete sources.
}
\end{figure}

\begin{figure}
\includegraphics{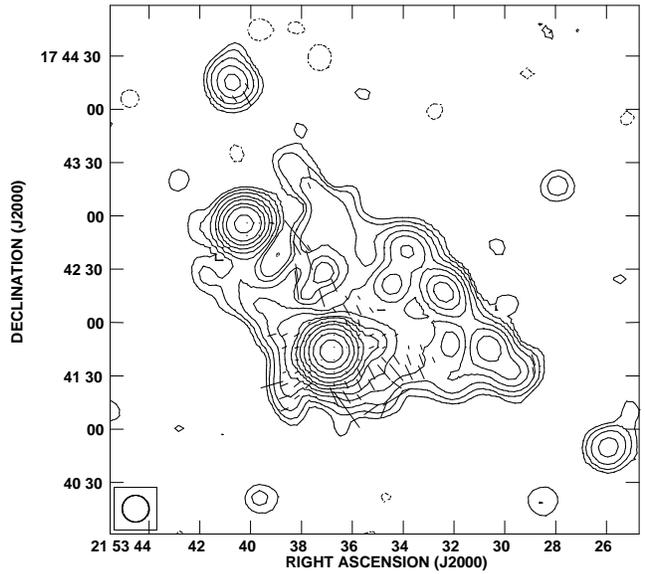}
\vspace{8 cm}
\caption{Contour map of the radio emission 
at 20 cm in A2390, with  superimposed lines representing
 the orientation of the polarization
vector (E-field).  Lines  are proportional in length to 
the fractional polarization: 1\arcsec~ corresponds to 2\%.
The image resolution and the 
contour levels are the same as in Fig. 9.
}
\end{figure}

\subsection{Abell 2390}

The diffuse emission in the center of this cluster is presented in
Fig. 10. It is quite irregular showing sharp and almost straight edges
toward south and east, and filaments to the north (note the hole of
emission north of the source S5).  The western edge is confused by
several discrete radio sources. The strongest source at the center,
labelled S5, is identified with the cluster dominant cD galaxy (Pierre
\etal 1996). This source is barely resolved on the arcsec scale (size
$\sim$ 3\arcsec~, Owen \etal 1993)

The analysis of optical and X-ray ROSAT data (Pierre \etal 1996,
B\"ohringer \etal 1998) indicates that the center of the thermal gas
distribution coincides with the position of the cD galaxy and that the
cluster is characterized by a massive cooling flow.
Recent Chandra data (Allen \etal 2001) confirm
the presence of a cooling flow in the cluster central region
and show in addition some substructure on the large scale 
(\gtsim~ 2\arcmin), suggesting 
that the cluster has not fully relaxed following its most recent
merger activity. 

The diffuse radio emission, located at the
cluster center and smaller than 2\arcmin~ in size, 
is unlikely to be related to the X-ray substructure, 
i.e. the cluster merger.
Based on the presence of a cluster cooling flow and of
a strong central radio galaxy, the diffuse source in this cluster
naturally falls in the class of mini-halos, i.e. halos of moderate
size around powerful radio galaxies at the center of cooling flows (as
in the Perseus cluster). 
The mini-halo in A2390 is similar in size to
that of the Perseus cluster ($\sim$ 640 kpc in diameter, scaled to our
Hubble constant, Burns \etal 1992),  but its structure is remarkably
non spherical and more irregular (see also the minihalo in A2626,
Gitti \etal 2002b).

This is the only diffuse source of the present sample where polarized
flux has been detected (see Fig. 11).  The polarization percentage is
highest in the southern region of the diffuse emission, with values
between 10\% and 20\%, whereas in north-western and eastern regions
the polarization degree is of $\sim$8-10\%.  The strong central source
identified with the cD galaxy is almost unpolarized ($<$1\%).  The
equipartition magnetic field is of the order of 1 $\mu$G as expected
from the model of Gitti \etal (2002a). The orientation of the magnetic
field, predicted to be isotropic by the above mentioned model, cannot
be established with 1.4 GHz data, which could be affected by Faraday
Rotation.

\begin{figure}
\includegraphics{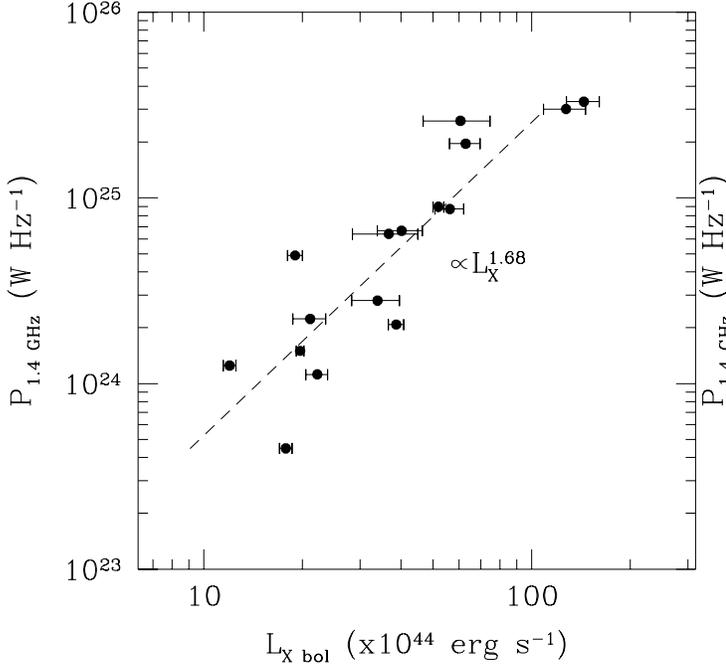}
\vspace{10 cm}
\caption{
Plot of the monochromatic radio power at
1.4 GHz of the halos larger than 1 Mpc versus the cluster
bolometric X-ray luminosity. The dashed line represent the 
best fit to the data.
}
\end{figure}

\begin{figure}
\includegraphics{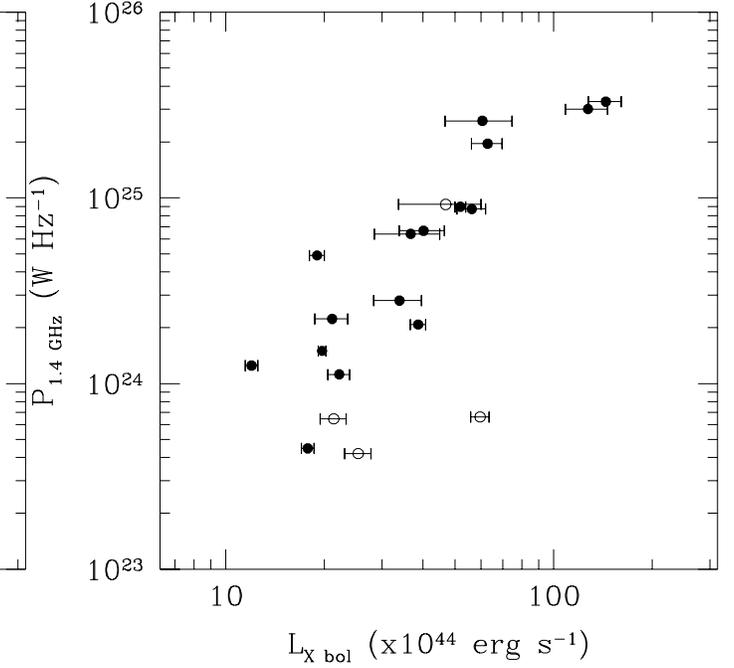}
\vspace{10 cm}
\caption{
Halo monochromatic radio power at
1.4 GHz versus the cluster bolometric X-ray luminosity for halo sizes
$>$1 Mpc ($\bullet$) and $<$ 1 Mpc ($\circ$).
}
\end{figure}

\section{Correlation between halo radio power and cluster X-ray luminosity}

The presence of a correlation between the halo radio power and the
cluster X-ray luminosity has been first noted by Liang \etal (2000)
and confirmed in further studies (Feretti 2000, Govoni \etal 2001).
By adding the present data to data in the literature, we obtain the
correlation in Fig. 12.

We stress here that the correlation applies to clusters showing major
merger events, and therefore cannot be generalized to all clusters.

Using only clusters with giant radio halos (size $>$ 1 Mpc), 
the best fit between radio and X-ray luminosity is 
P$_{\rm 1.4 GHz} \propto L_{\rm X}^{1.68\pm0.15}$. 
The addition of the smaller size halos (Fig. 13), add a
larger dispersion of the X-ray data in the region of lower radio
power.  The overall correlation does not change
significantly, however we note that at P$_{\rm 1.4 GHz}$ lower
than $\sim$ 3 10$^{24}$ W Hz$^{-1}$, 
the correlation seems to be virtually absent,
with the only indication that halos are present in clusters with X-ray
luminosity $>$10$^{45}$ erg s$^{-1}$.
No firm conclusion can be drawn, as the number 
of low power radio halos is presently quite small.

An extrapolation of the large radio halo correlation to low radio and
X-ray luminosities indicates that clusters with L$_{\rm X}$
\ltsim~10$^{45}$ erg s$^{-1}$ would host halos of power of a few
10$^{23}$ W Hz$^{-1}$.  With a typical size of 1 Mpc, they would have
a radio surface brightness lower than current limits obtained in the
literature and in the NRAO VLA Sky Survey.  Therefore, it is possible
that future new generation instruments (LOFAR, SKA) will allow the
detection of low brightness/low power large halos in virtually all the
merging clusters.  On the other hand, future highly sensitive data
will clarify whether the correlation holds also at low power and for
small size halos, or the cutoff in the X-ray luminosity suggested by
Fig. 13 is real.

\section{Conclusions}

VLA images of 6 clusters of galaxies confirm the presence of a central diffuse
radio halo source in A401, A545, A754, A1914, and A2219 while in A2390 a 
mini-halo source has been found.

In A754 a second diffuse
component has been detected. 
The two diffuse sources in this cluster could be classified as a halo
and a relic source, but we cannot exclude that they are actually
two halos or two
relics, given the very complex structure of this cluster.

The radio halo in A2219 is among the largest sources known in the
literature: it is 10 times more powerful than the halo in the Coma
cluster and more than 2 Mpc in size.

Present results confirm the correlation of radio halos with cluster
merger processes while mini-halos seem to be correlated to the presence
of an active cD galaxy in cooling flow clusters.

Adding these new data to the literature data we obtained a sample of
16 well known radio halos with a linear size larger than 1 Mpc. For these
sources we derived the correlation between the radio and X-ray luminosity.
The estimated best fit is P$_{\rm 1.4 GHz} \propto
L_{\rm X}^{1.68\pm0.15}$, suggesting that clusters with a low X-ray 
luminosity ($<$ 10$^{45}$  erg s$^{-1}$) would host
halos too faint to be detected with the present generation of radio telescopes.
However, the addition of smaller size radio halos 
could suggest the presence of a cutoff in the previous 
correlation:
radio halos could be absent in clusters with a X-ray luminosity lower than 
\ltsim10$^{45}$ erg s$^{-1}$.
The future generation of new radio telescopes (LOFAR, SKA) will allow the
investigation of this point.

\begin{acknowledgements}

We thank the referee David Buote for his useful
comments, and for pointing out the X-ray structure of A1914,
as detected from the archive Chandra ACIS image.
This work was partially funded by the Italian Space Agency.

\end{acknowledgements}

\end{document}